# Postural adjustments preceding string release in trained archers

Authors: A. Kuch[1,2], R. Tisserand[2,3], F. Durand[4], T. Monnet[2], J-F. Debril[4]

[1]CRITT Sport Loisirs, Châtellerault, France
[2]Institut PPrime (UPR 3346), Université de Poitiers-ISAE-ENSMA, CNRS, Chasseneuil-du-Poitou, France
[3]Centre de Recherches sur la Cognition et l'Apprentissage (UMR 7295), Université de Poitiers, Université de Tours, CNRS, Poitiers, France
[4]CAIPS-CREPS de Poitiers, Vouneuil-sous-Biard, France

# Abstract

Optimal postural stability is required to perform in archery. Since the dynamic consequences of the string release may disturb the archer's postural equilibrium, they should have integrated them in their motor program to optimize postural stability. This study aimed to characterize the postural strategy archers use to limit the potentially detrimental impact of the bow release on their postural stability and identify characteristics that may explain a better performance. Six elite and seven sub-elite archers performed a series of 18 shots at 70 meters, standing on two force plates. Postural stability indicators were computed during the aiming and the shooting phase using the trajectory of the center of pressure. Two postural strategies were defined, as whether they were triggered before (early) or after (late) the string release time. Both groups used anticipated postural adjustments, but elite archers triggered them before the string release more often and sooner. Scores differed between the two groups, but no differences were found between early and late shots. Trained archers seem to have finely integrated the dynamic consequences of their bow motion, triggering anticipated postural adjustments prior to the string release. However, it remains unclear whether this anticipation can positively influence the performance outcome.

# Highlights

- The current study focuses on postural control of highly trained archers during the aiming and shooting phases.
- Because the string release is voluntary triggered, archers use anticipatory postural adjustments to reduce its consequences on their postural equilibrium.
- The main difference between the postural adjustment performed during anticipated and non-anticipated shots is the release timing onset.
- Anticipating the string release postural perturbation indicate archers have finely integrated its dynamic consequences in their motor program but it does not seem to influence the performance outcome.

# 1.Introduction

Olympic archery is a sport requiring to shoot arrows with a recurve bow at a target located 70 meters away [1]. To achieve both accuracy and precision, elite archers are trained to stand upright as quietly as possible while aiming at the target. Indeed, natural whole-body oscillations observed during standing prior to string release may result in aiming error and therefore reduce the shooting performance [2]. But precisely shooting arrows is not only about aiming. When archers release their fingers, the string propels the arrow in the direction of the target with less than 16 ms of string and arrow contact [3]. For modern bows, the recoil kinetic energy is estimated at around 0.2 J, which leads to a recoil velocity of about 250 mm/s [4]. This bow recoil can potentially disturb archers' postural equilibrium and contribute to poorer shooting performance [5]. Yet, how elite archers deal with the consequences of this voluntary triggered movement and whether or not their postural strategy has an impact on their shooting performance remains unclear.

Postural control serves two main purposes: orientation and balance [6]. When all forces applied on a body are balanced, it can be assumed that its postural balance is regulated by the displacement of the center of pressure (CoP) to maintain the vertical projection of the center of mass (CoM) within the base of support (BoS) [7]. According to the simple and widely used inverted pendulum model [7], when postural balance is challenged by motion of the CoM in one direction, the CoP has to move in the same direction and with larger amplitude to keep the CoM projection within the BoS. When the CoM motion comes from a self-generated voluntary movement, i.e. an 'internal' perturbation, the postural control strategy is thought to be pro-active [8,9]. This means that the dynamic consequences of the forthcoming perturbation are predicted and therefore integrated in the following motor response plan.

Thus, the dynamic consequences of the movement perturbation on posture can be minimized through anticipatory postural adjustments (APA) [8,10,11], starting before the movement onset [12], to maintain both postural orientation and balance [13].

In archery, a shot is generally divided into two main phases: i) an aiming phase, where the archer aims at the target while maintaining an upright quiet standing posture, and ii) a shooting phase, where the string is released and the arrow flies toward the target [1]. These phases are separated by the clicker fall. The clicker is a metallic device mounted on the bow riser to optimize bow draw length consistency [1]. Its fall emits a sound, signaling to the archer that the correct bow draw length has been reached. Archers are taught to release the string only after hearing this sound, and a shorter duration between the clicker fall and the string release was linked to increased performance [14]. Like reacting to the clicker fall, postural stability is a feature that can be trained and improved through regular practice [15,16]. In standing shooting disciplines like archery [14,17–20], rifle [21–24] or biathlon (standing shooting) [25], analysis of postural balance during the aiming phase showed that better performance in skilled shooters compared to less skilled counterparts could be explained by smaller postural sway characteristics. However, very few studies characterized postural balance during the shooting phase, leaving the question of how skilled shooter minimize the dynamic consequences of their shooting activity on postural stability partially unanswered. In air pistol shooting, electromyographic analysis revealed the presence of APA specific to the direction of the movement and muscular activity starting up to 100 ms before the shooting onset [26]. Yet, their mechanical contribution to postural balance have never been characterized in archery.

The goal of our study is to characterize the postural strategy used by trained archers to limit the potentially detrimental effect of the string release on postural balance and determine whether or not this has a positive influence on performance. Since trained archers release their string voluntarily, we first hypothesized that APA will be triggered in the direction of the bow recoil, to limit its dynamic impact on postural balance, and that these APA will be finely timed with the clicker fall. Second, since APA can be trained to limit the detrimental consequences of self-generated perturbation on postural balance [16], we hypothesized that APA characteristics will be mechanically more efficient in elite archers than in sub-elite archers, showing a better dynamic postural control that could partially explain performance difference between the two levels of expertise.

# 2. Material and Methods

## 2.1. Participants

Thirteen archers participated in this study. Six participants (3 males, 3 females, age 27.0 ± 4.7, height 174.3 ± 8.2 cm, mass 80.6 ± 27.2 kg) were recruited in the French National elite training center and were placed in the elite group (16.0 ± 1.9 years of practice), according to the following inclusion criteria: over 10 years of archery practice and at least one participation to a senior international outdoor archery competition (European Championships, World Cup stages, World Championships, Olympic Qualifiers or Olympic Games) in the year prior to the experiment. Seven participants (4 males, 3 females, age 16.5 ± 4.9, 169.2 ± 3.6 cm, mass 59.7 ± 4.8 kg) were recruited in a next-generation training center and were placed in the sub-elite group (8.0 ± 1.8 years of practice), according to the following inclusion criteria: less than 10 years of archery practice and never competed at a senior international outdoor archery competition. For both populations, the exclusion criteria was to not have suffered of any injury 6 months prior to the experiments. The experiment was approved by the department local ethics committee and conducted with the informed and written consent of all participants.

2.2. Experimental design

Each participant first completed their usual competitive warm-up routine for 15 minutes. Warm-up routines included exercises mobilizing upper limbs joints and muscles followed by taking a dozen of shots at a standard 10-ring competition target face of 122 cm diameter, located at the Olympic distance of 70 meters. During the experiment, each participant was asked to shoot 18 arrows on that same target, in the same conditions (no wind). Participants executed every shot using their own recurve bow, equipped with stabilizers and sight, as well as their own arrows. Participants started every trial with both feet on the ground, wearing shoes. Before taking a shot, participants stepped on two force plates, by placing each foot on one of the two (Figure 1A). When installed, participants could shoot one arrow at their own pace. Once the arrow reached the target, participants stepped out of the force plate and started over. The score of each arrow was collected.

2.3. Instrumentation

To compute the CoP before and during the shot, we used the ground reaction forces and torques from two force plates (600x500x50 mm, Kistler 9260AA6, Winterthur, Switzerland) sampled at 1000 Hz. The force plates were placed next to each other and oriented so that the medio-lateral axis (ML) was aligned with the shooting direction and the antero-posterior axis (AP) was the axis orthogonal to the shooting direction (Figure 1A). Signals from the force plates were retrieved using Bioware software (Kistler 2812A, version 5.3.2.9), filtered with a 3$^{rd}$ order Butterworth low-pass filter with a 6 Hz cutoff frequency [14] and used to process the global CoP coordinates according to Kistler recommendations.

**FIGURE 1 ABOUT HERE**

To identify characteristic instants relative to the different phases in archery, we filmed every shot with a high-speed camera (Photron FASTCAM Mini UX100 type 800K-C-16G, Tokyo, Japan) sampled at 1000 Hz. The camera was placed 2 m away from the participant, filming the bow riser on the clicker's side (Figure 1B). Force plates and the high-speed camera were synchronized using a BNC trigger port on the data acquisition system (Kistler DAQ 5691A, Winterthur, Switzerland). Each measurement started with an external wired trigger button pressed by an experimenter.

2.4. Postural stability

2.4.1 Temporal identification

Images from the high-speed camera were used to identify two decisive instants: the clicker falling time and the string release time. Here, the clicker falling time was identified as the first image where the clicker touched the bow riser (Figure 1B right), and the string release time was identified as the first image where the arrow tip moved towards the target. To compute each participant's reaction time, the clicker falling time was subtracted to the string release time. Postural stability of participants during the aiming phase was characterized on a period of 1 second prior to the clicker falling time [14]. This time interval was chosen as a minimum duration prior to the clicker fall observed for all the participants between aiming at full draw and releasing. The shooting phase was considered to be the period starting after to the clicker falling time until 0.5 second after the string release time.

2.4.2 Postural stability during the aiming phase

During the aiming phase, six postural stability indicators were computed [27]: the mean ($d_{mean}$) and root mean square distance ($d_{rms}$) around the mean CoP, the mean ($v_{mean}$) and maximum velocity ($v_{max}$) of the CoP, in both the ML and AP axis as well as in the $Oxy$ plane;

the sway amplitude ($A$) only in both ML and AP directions; and finally the 95% confidence ellipse area ($A_{95\%}$) in the $Oxy$ plane.

2.4.3 Postural stability during the shooting phase

A CoP displacement away from the target was observed in the ML shooting axis in every trial (Figure 2). This ML CoP movement onset ($t_0$) was computed as the instant from the clicker falling time when the participant's sway velocity in the ML axis crossed a threshold set at two standard deviations above $v_{mean}$ measured during the aiming phase. To characterize the postural strategy during the shooting phase, we assessed the CoP trajectory on a period of 0.5 second after $t_0$.

For analysis purposes, we used the timing of $t_0$ to characterize the archers' postural strategy. If $t_0$ occurred prior to the string release which constitutes the perturbation, the postural strategy was labelled as "early", otherwise it was labelled as "late". Usage of either one or the other strategy was quantified in each group.

2.5. Statistical analysis

A total of 230 shots (elite $n = 107$, sub-elite $n = 123$) were successfully recorded and analyzed. Normality of the different dependent variables was tested with a Shapiro-Wilk test. Since those were not normally distributed, results are presented with medians and interquartile ranges unless said otherwise. For all the CoP indicators measured during the aiming phase, two variables were extracted to account for intra-individual variability. First, results of all 18 trials from one archer were averaged and then averaged among all the archers from a same group (elite $n = 6$, sub-elite $n = 7$). Second, the standard deviation of all 18 trials for one archer was computed, followed by the standard deviation of all archers from a same group. These two results (average and standard deviation) were then compared for each CoP indicator using a non-parametric Mann-Whitney U test between the two groups.

Reaction time and $t_0$ were compared using Kruskall-Wallis tests with two factors (Group: elite vs sub-elite and Strategy: early or late), with Dunn-Sidak tests as post-hoc. Differences in percentage of use of the early or the late strategy between the two groups were evaluated using a Fisher's exact test.

Since every archer had a different distribution of early and late shots, to avoid any individual influence on the group results, the CoP trajectories during the shooting phase were averaged to form one continuous timeseries representative of each archer (elite $n = 6$, sub-elite $n = 7$) in each strategy. Then, one dimensional statistical parametric mapping (1D SPM) [28], using non-parametric two sample t-tests [29], was conducted to analyze the influence of expertise within strategies and strategy within groups, and to identify any differences in the continuous ML CoP trajectory after $t_0$.

Score distributions were not normal, according to a Shapiro-Wilk test. Therefore, non-parametric Mann-Whitney U test was used to compare score performance between both groups and both postural strategies. Additionally, a Kruskal-Wallis test was performed to compare the effect of the strategy on the performance score for each group, with Dunn's test used as post-hoc with a Holm-Bonferroni correction. For all statistical tests, a $p < 0.05$ was set for significance. Data processing was performed using customized routines written in Matlab (version R2021b, The MathWorks, Natick, USA) and the open source spm1d Matlab package[1] (version M.0.4.8). Statistical analysis was conducted using the open source statistical software JASP (version 0.16.4, Amsterdam, Netherlands).

# 3. Results

---

[1] https://spm1d.org

3.1 Postural stability during the aiming phase

Averaged CoP indicators and their respective variability calculated during the aiming phase are presented in Table 1 with the results of their statistical comparison between the two groups. For the averaged results no differences were found between the two groups, except for $d_{mean}$ and $d_{rms}$ that were smaller in the elite group compared to the sub-elite group in both the AP axis and the $0xy$ plane, as well as $A\ AP$ (p < 0.05). For the variability results no differences were found between the two groups, except for $v_{max}$ that was larger in the elite group compared to the sub-elite group in both the ML axis and the $0xy$ plane (p < 0.05).

**TABLE 1 ABOUT HERE**

3.2 Postural stability during the shooting phase

Kruskal-Wallis test indicated evidence of differences between the pairs of group/strategy for reaction time (H(3) = 28.1, $p < 0.001$) and $t_0$ (H(3) = 174.9, $p < 0.001$). The elite group had a shorter reaction time compared to the sub-elite group (146 ± 12 ms versus 156 ± 15 ms, $p < 0.001$). Within strategies, reaction time was shorter in the elite group compared to the sub-elite group for the early strategy only (145 ± 11 ms versus 157 ± 14 ms, $p = 0.02$). Within groups, no difference was found for the $RT$ between early and late strategies (both $p > 0.05$). Within groups, $t_0$ occurred significantly sooner in the early than in the late strategy for both the elite (23 ± 38 ms versus 163 ± 24 ms, $p < 0.001$) and the sub-elite archers (119 ± 37 ms versus 180 ± 26 ms, $p < 0.001$). Within strategies, $t_0$ occurred sooner in the elite group compared to the sub-elite group for the early strategy only ($p < 0.001$).

**FIGURE 2 ABOUT HERE**

The early and late strategy were used by every archer regardless of their level of expertise. However elite archers used the early strategy more often than the sub-elite group (82 trials out of 107 or 77 ± 16 % versus 73 trials out of 123 or 59 ± 17 % of early shots, respectively, $p = 0.007$).

The 1D SPM analysis (Figure 3) showed no difference among groups or strategies for the CoP trajectory in the ML axis for a period of 0.5 seconds after $t_0$ ($p > 0.05$).

3.3 Scores

The elite group scored higher (9.2 ± 0.8 (mean ± SD)) than the sub-elite group (8.9 ± 1.0), $p = 0.03$. However, no differences were found for scores between the trials where an early strategy was used (9.0 ± 0.8) and where a late strategy was used (8.9 ± 1.1) nor when scores of early and late strategies were compared within groups (all $p > 0.05$).

**FIGURE 3 ABOUT HERE**

# 4. Discussion

The objective of this study was to characterize the postural strategy used by trained archers to limit the potentially detrimental effect of string release on postural balance and determine whether this strategy could explain differences in performance outcome between two levels of expertise. As expected, all archers displayed APA in the shooting direction and with greater amplitude than the balance perturbation. Moreover, these APA were finely timed clicker fall, especially in elite archers but with no evident impact on scoring performance. These results suggest that the mechanical consequences of the bow recoil are finely integrated into the motor program released for the shooting phase.

4.1 Postural stability during the aiming phase

During the aiming phase, three averaged postural sway indicators ($d_{mean}$, $d_{rms}$ and $A\ AP$) were smaller in the elite group compared to the sub-elite group for the AP component (Table 1). The variability analysis showed there was no difference between groups for those three indicators (Table 1), confirming the difference in the averaged results is not due to intra-individual variance. This suggests that the determinant axis for postural stability during aiming in trained archers is not the shooting direction axis but the one perpendicular to it. In a bipedal standing posture, the BoS is generally smaller in the AP than in the ML direction and actions of the mono-articular ankle muscles, such as the soleus and tibialis anterior, are mainly responsible for accurate movement of the CoP in the AP direction [7]. In the ML direction, stronger hip muscles are thought to be responsible for CoP sway control [7]. This can explain that even in young adults, larger postural sway characteristics are found in the AP axis than in the ML axis in quiet standing postures [27]. Thus, fine postural balance control for elite archers seems to be principally controlled in the AP direction [24,25]. The difference between the elite group and the sub-elite group corroborates previous results showing that reduced CoP sway during aiming is positively correlated with expertise and performance in archery [14,17–20], akin to other shooting disciplines [21–25]. Along with these studies, we support the view that elite archers perform better than non-elite archers partly because they reduce the AP oscillation of their CoP during the aiming phase.

Only one indicator (ML component of $v_{max}$) showed a larger variability among archers from the elite group compared to the sub-elite group (Table 1). This increased variability can be explained by the onset distribution of the ML CoP movement of elite archers (t0) that sometimes occurred before the clicker falling time (Figure 2) and may have been caught as the maximal CoP velocity of the aiming phase in our analysis.

4.2 Postural stability during the shooting phase

4.2.1. An anticipated postural strategy finely integrating the mechanical consequences of the string release

Releasing the string frees the potential energy accumulated in the bow limbs, inducing a recoil movement of the bow [4]. From a bow recoil velocity estimated at 0.26 m/s [4], a string and arrow common motion time of 16 ms during the internal ballistic phase [3] and assuming a rigid body system including both the archer and the bow, we estimated the displacement of the system CoM at approximately 4 mm in the direction opposite to the target. According to the inverted pendulum model, widely used to study the mechanics of postural control [7], this recoil caused by the recoil motion of the bow may threaten the mechanical balance conditions achieved during the aiming phase. To keep balancing the pendulum, a CoP movement towards the back leg is necessary to slow down the CoM motion and restore postural balance. For all participants and in all trials, we observed that the CoP indeed moves away from the target, with an amplitude between 15 to 20 mm, which is about 8 to 10 times larger than oscillations measured during the aiming phase (1.9 mm) and 4 to 5 times larger than the estimated CoM displacement caused by the bow recoil perturbation. Besides, the ML CoP sway amplitude remains consistent within groups and strategy (Figure 2 & Figure 3). Because our participants were all trained archers, they are used to experience the mechanical consequences of string release and therefore seem able to generate APA finely matching the spatial characteristics of an expected postural perturbation [30,31].

The string release is extremely short (< 16 ms of string and arrow contact [3]). Even though the CoP is assumed to move without inertia, its motion depends on the contraction of postural muscles that have a certain latency [7]. Therefore, lower-limb muscle contractions should be initiated before the shooting onset [26] so that the CoP can be closer to the maximum recoil amplitude at the string release time (between 30 and 66% here, Figure 2). In a large majority of trials (67% for all participants), $t_0$ preceded the perturbation and occurred only 23 ms after the clicker fall on average in elite archers, which is below any postural reflex latency response (~40 ms [32]). The distribution of $t_0$ were also quite small in both groups (< 40 ms). Together, these results suggest motor commands responsible for the CoP recoil motion were pre-programmed beforehand [8,9] and released at a similar dedicated timing. By initiating a large CoP movement preceding the perturbation, our study suggests that trained archers attempt to release pre-programmed anticipatory postural commands that aim to counteract the mechanical consequences of the string release to maintain balance. The high reproducibility in the timing release of this postural strategy also suggests that archers have finely integrated [33] the timing of the string release in their shooting motor commands.

4.2.2. Expertise makes archers trigger their postural strategy sooner

When the temporal delay of $t_0$ was removed, both groups exhibited the same postural response in terms of ML CoP movement (Figure 3). This result suggests that sub-elite archers have integrated the spatial characteristics of the bow recoil perturbation in their motor program. Both groups of archers did trigger the string release after hearing the clicker fall sound in the range of reaction times following expected auditory stimuli (120-160 ms [34]). However, the release of the APA did not seem to be timed with the same event in the two groups. Elite archers seemed to release their postural strategy at two distinct temporal instants (Figure 2). The shots labelled as early were released 23 ms ± 38 ms after the clicker fall, which is about 130 ms before the string release; whereas the shots labelled as late were released 163 ± 24 ms after the clicker fall, which is about only 15 ms after the string release. With 77 ± 16 % of early shots, our results suggest elite archers aim to release their postural strategy with the clicker fall. On the contrary, sub-elite archers seemed to release the same postural strategy in terms of amplitude, but closer to the string release only. In the latter group, the shots labelled as early were released 119 ms ± 37 ms after the clicker fall, which is about 30 ms before the string release; whereas the shots labelled as late were released 180 ± 26 ms after the clicker fall, which is about 30 ms after the string release. The timing between the ML CoP movement onset in early shots and string release is much shorter than APA measured in predictable perturbation condition (100-150 ms range [35]) and its proximity to the late shots (60ms) point toward an anticipating postural strategy timed on one event, suggesting sub-elite archers are anticipating the string release event more than the clicker fall. This observation questions our arbitrary choice to use the perturbation as a temporal threshold to categorize early and late shots, which seems less suitable for the analysis of postural strategy in sub-elite archers than in elite archers (Figure 2). In the same way that APA are developed and improved through extensive training in sports requiring enhanced postural control such as slacklining or gymnastics [15], our findings suggest that elite archers have updated their internal models of postural control [36] whereas adolescent sub-elite archers may suffer from sensorimotor representation that are not completely mature [37,38] due to shorter experience [15] and/or biological maturation not fully completed [39].

As expected, the performance score was higher for the elite group compared to the sub-elite group [20], but scores between the early and late strategies within these groups did not show any statistical differences, despite being slightly higher for the early strategy. Therefore, an anticipatory postural strategy released sooner does not seem to have an influence in the performance score. However, in this study the participants shot only 18 arrows which is not sufficient to ascertain if those postural strategies differences would indeed affect the global score in competition [40]. In a world elite event archer shoot 72 arrows just in the qualification round. According to a model proposed of losing 1.4 points per ranking place [41], such slightly lower score per arrow (0.2 points) can translate into the loss of 10 ranking places, which would harden the matchup for the face-off rounds.

### 4.3. Limitations

Studying elite level athletes comes with inherent difficulties, such as the limited number of participants and their tight schedule compatibility. Only trained archers participated in this study, which may limit the generalization of our results to all archers. Measuring the postural stability of the same group of young archers years later after full maturation of their body schema and/or biological maturation, as well as in beginner archers may better clarify how the postural strategy described here is learned and can be trained. Electromyography measurements, often used in anticipated postural adjustments studies [10,11,13] were not considered in order to conduct the experiment without disturbing the archer's shooting sequence. Further experiments with muscular activity monitoring devices could pinpoint which muscles are pre-activated prior to the string release and complement the understanding of this anticipatory strategy.

### 4.4. Conclusion

By releasing pre-programmed anticipatory postural adjustments at a desired instant during their shooting sequence, trained archers seem to have fully integrated the characteristics of the balance disturbance of the string release on their whole-body upright posture. Elite archers seem to time the releasing of this anticipatory strategy with the clicker fall, whereas younger sub-elite archers time the releasing of this strategy rather with the string release instant. However, this difference in timing between the two levels of expertise did not allow to conclude as to whether releasing this strategy sooner helps increases performance.

# Declaration of interest statement

# Tables

Table 1 Postural stability indicators computed during the aiming phase and compared between the two groups. Abbreviations stand for: mean ($d_{mean}$) and root mean square distance ($d_{rms}$) around the mean CoP, the mean ($v_{mean}$) and maximum velocity ($v_{max}$) of the CoP, the sway amplitude ($A$) and the 95% confidence ellipse area ($A_{95\%}$). ML = mediolateral (shooting) direction. AP = anteroposterior direction. *$p < 0.05$

| | | Averaged indicators | | | Variability indicators | | |
|---|---|---|---|---|---|---|---|
| | | elite (n = 6) | sub-elite (n = 7) | | elite (n = 6) | sub-elite (n = 7) | |
| | | median (IQR) | median (IQR) | $p$ | median (IQR) | median (IQR) | $p$ |
| ML Direction | $d_{mean}$ ML (mm) | 0.41 (0.15) | 0.43 (0.22) | 0.73 | 0.17 (0.01) | 0.14 (0.09) | 0.30 |
| | $d_{rms}$ ML (mm) | 0.49 (0.18) | 0.51 (0.28) | 0.63 | 0.19 (0.03) | 0.17 (0.11) | 0.45 |
| | $v_{mean}$ ML (mm/s) | 6.84 (3.85) | 7.52 (5.2) | 0.84 | 1.51 (1.01) | 1.3 (1.04) | 0.84 |
| | $v_{max}$ ML (mm/s) | 32.54 (13.85) | 23.66 (16.75) | 0.30 | 24.13 (8.73) | 4.22 (4.37) | **0.01*** |
| | $A$ ML (mm) | 1.98 (0.85) | 1.95 (1.2) | 0.95 | 0.86 (0.20) | 0.65 (0.58) | 0.37 |
| AP Direction | $d_{mean}$ AP (mm) | 0.70 (0.12) | 0.86 (0.14) | **0.04*** | 0.46 (0.11) | 0.48 (0.10) | 0.37 |
| | $d_{rms}$ AP (mm) | 0.83 (0.12) | 1.03 (0.16) | **0.04*** | 0.53 (0.12) | 0.52 (0.14) | 0.53 |
| | $v_{mean}$ AP (mm/s) | 5.04 (0.49) | 6.52 (2.16) | 0.18 | 1.58 (0.47) | 1.47 (0.56) | 0.98 |
| | $v_{max}$ AP (mm/s) | 15.18 (2.43) | 20.13 (7.86) | 0.23 | 5.36 (1.84) | 4.62 (2.92) | 0.73 |
| | $A$ AP (mm) | 2.72 (0.41) | 3.60 (0.40) | **0.04*** | 1.64 (0.36) | 1.38 (0.33) | 0.99 |
| $Oxy$ plane | $d_{mean}$ (mm) | 0.94 (0.10) | 1.13 (0.07) | **0.04*** | 0.43 (0.14) | 0.44 (0.13) | 0.53 |
| | $d_{rms}$ (mm) | 1.04 (0.10) | 1.25 (0.07) | **0.02*** | 0.48 (0.16) | 0.47 (0.14) | 0.63 |
| | $v_{mean}$ (mm/s) | 9.80 (3.16) | 11.08 (6.11) | 0.63 | 2.14 (0.91) | 2.1 (1.29) | 0.95 |
| | $v_{max}$ (mm/s) | 27.07 (10.69) | 20.56 (14.09) | 0.37 | 23.94 (4.14) | 4.99 (2.39) | **0.01*** |
| | $A_{95\%}$ (mm²) | 7.59 (3.01) | 8.90 (4.09) | 0.23 | 5.85 (1.67) | 4.92 (1.81) | 0.63 |

18 # Figure 1

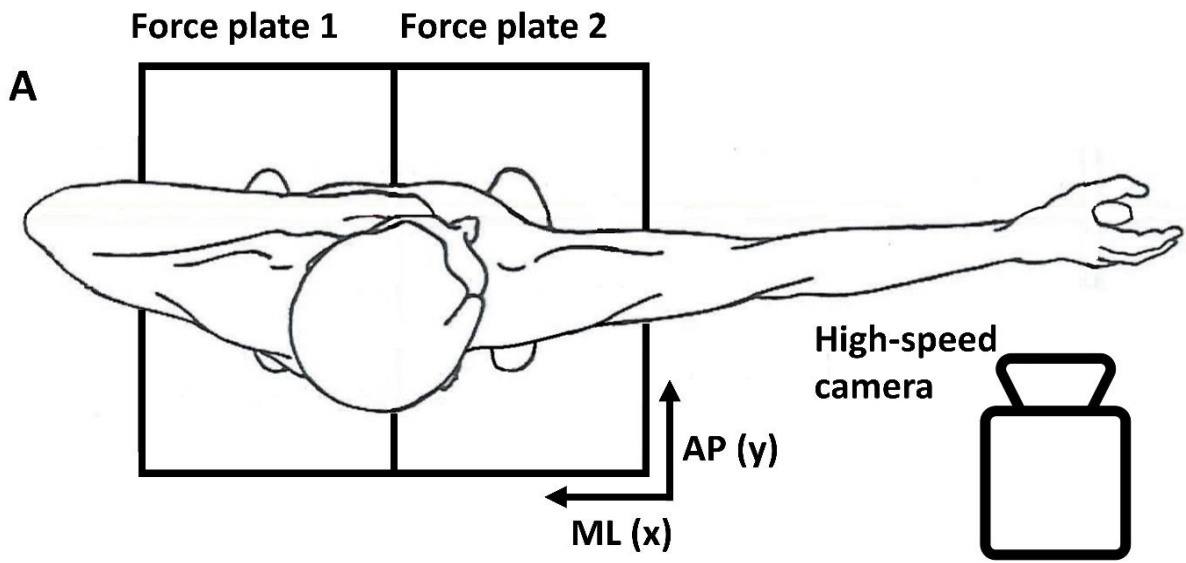

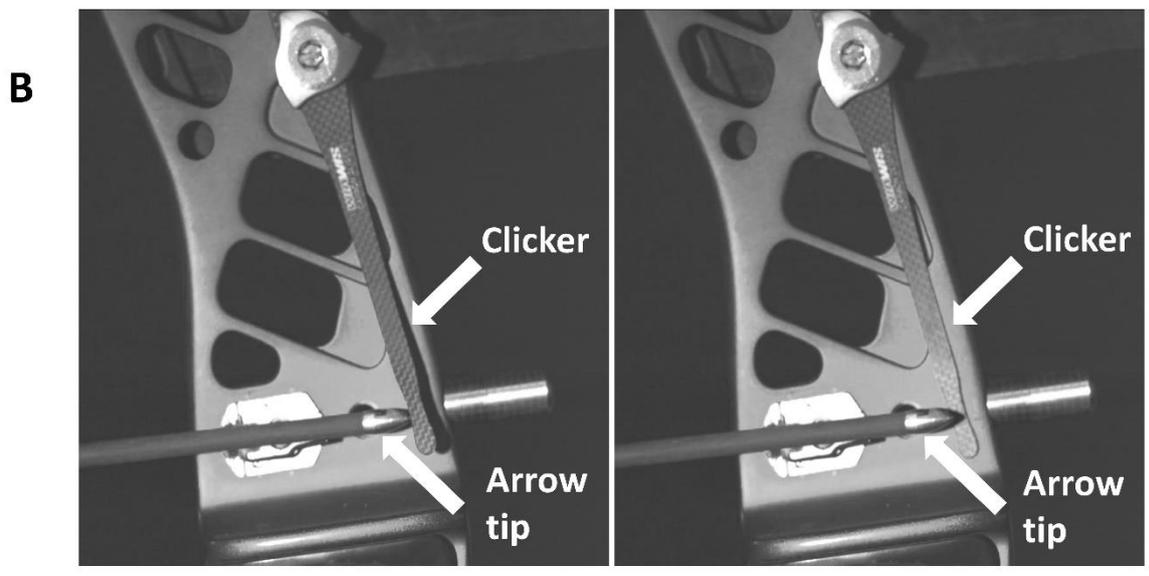

19
20

## Figure 2

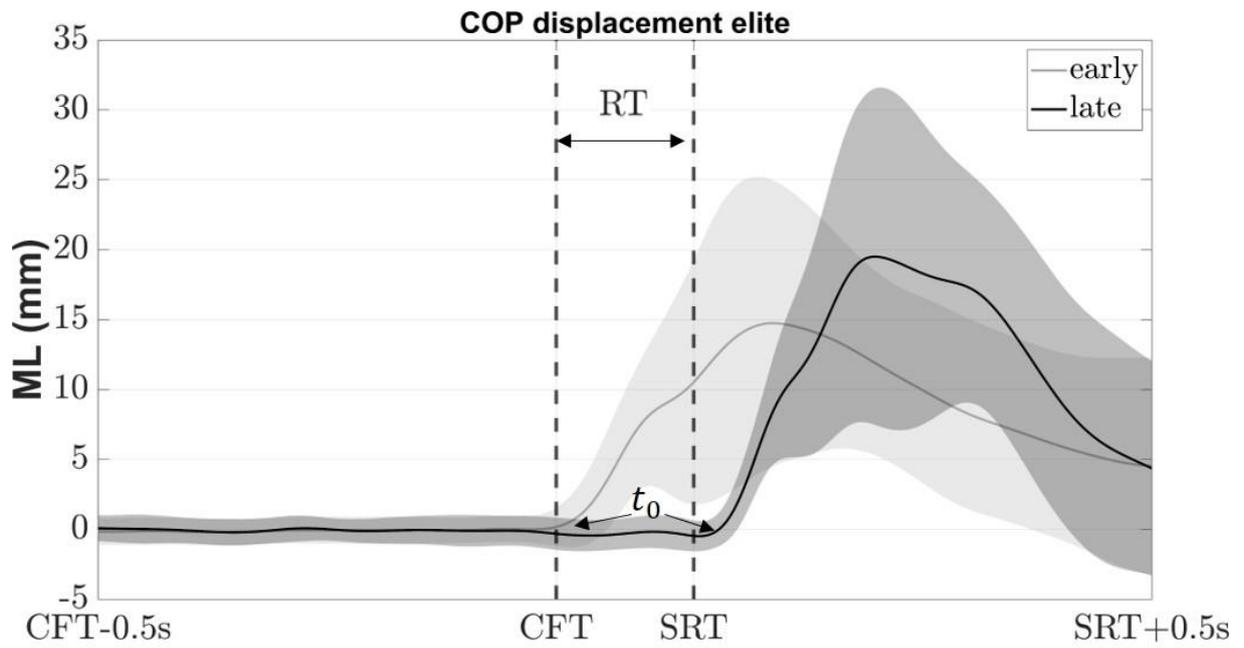

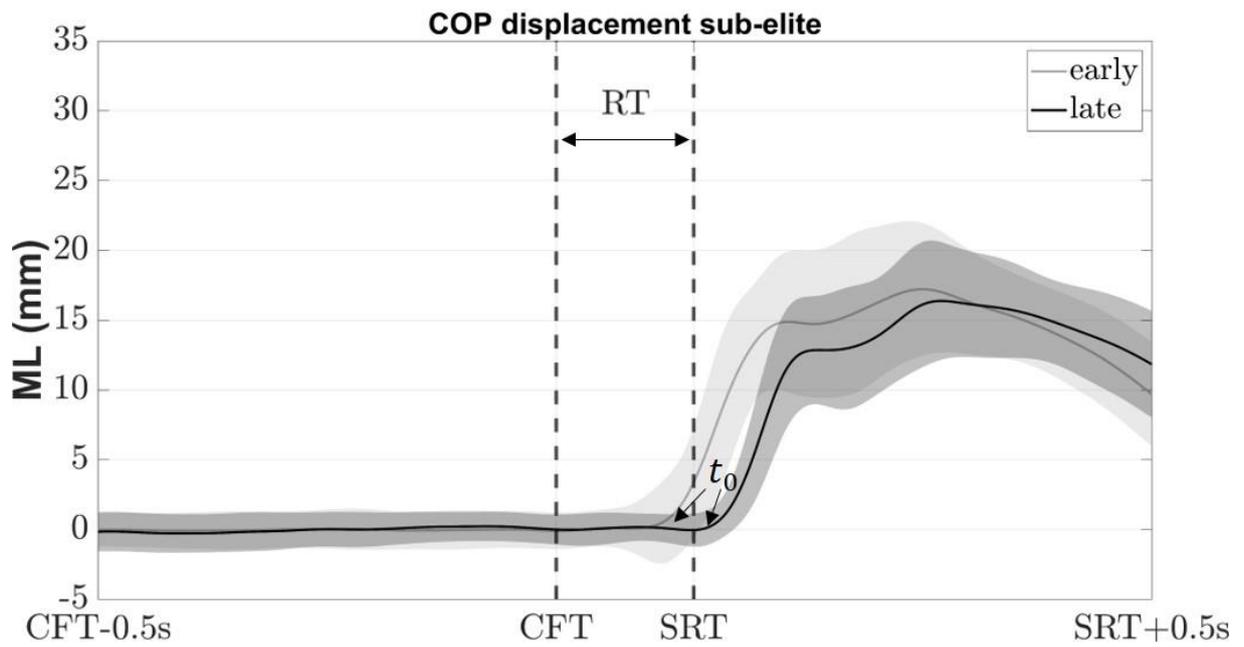

24  **Figure 3**

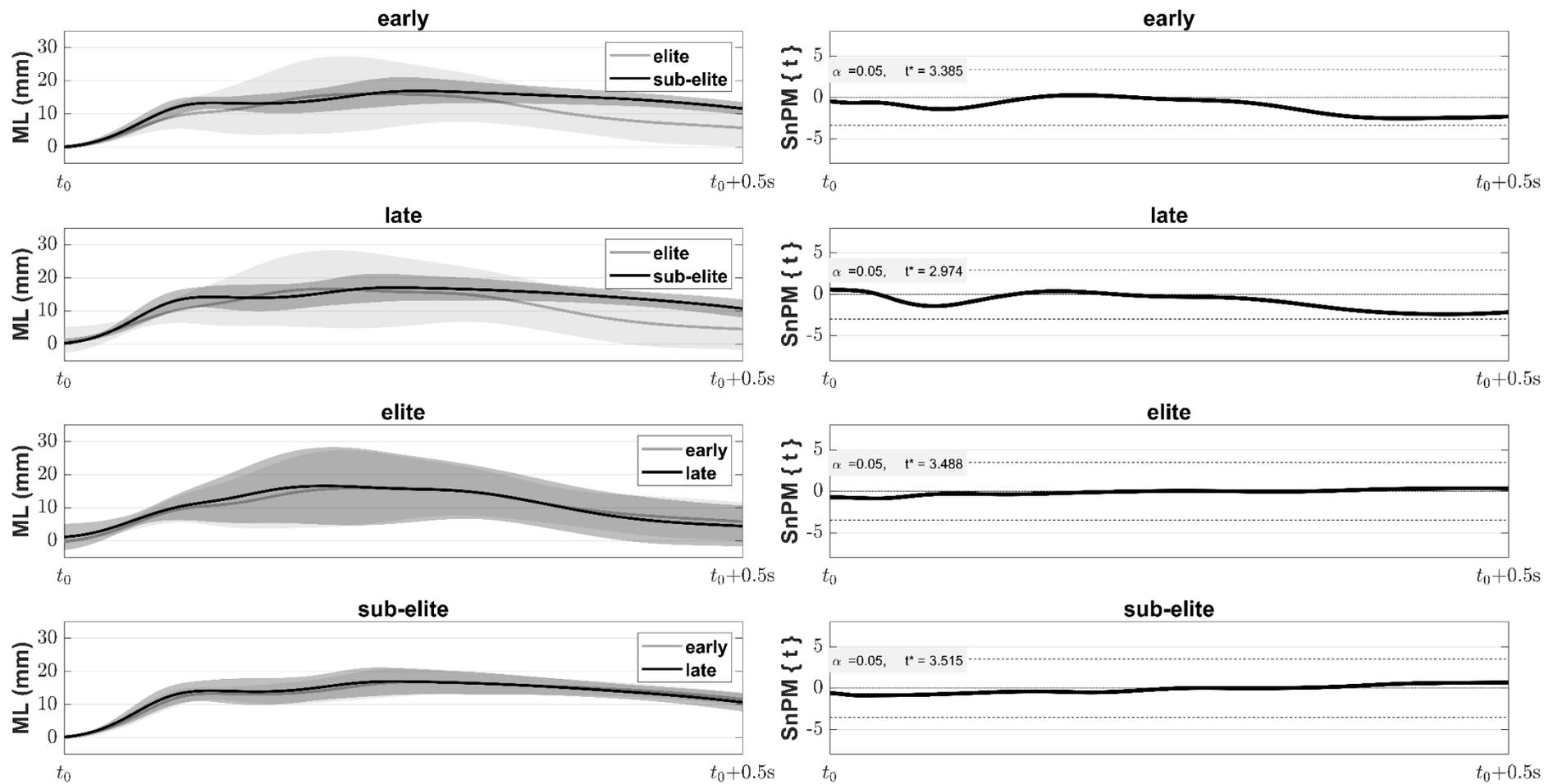

25
26

# Figure Captions

Figure 1. **A**. Reference frame, top view for a left-handed archer, with instrumentation used in the experiment (force plates, high-speed camera). The ML axis is defined as the shooting axis. **B**. Snapshots from the high-speed camera. Left panel: right before $CFT$ (clicker falling time), the arrow is almost at full draw behind the clicker. Right panel: $CFT$ instant, the arrow is at full draw and the clicker falls, giving an auditory stimulus that archers are taught to react to.

Figure 2. CoP timeseries in the ML axis (shooting direction) for both groups. Mean (plain lines) and standard deviation (shaded area) of all trials are presented for one postural strategy (early – light grey – or late – dark grey). For clarity purposes, the timeseries are presented from 0.5 second prior to the $CFT$ to 0.5 seconds after the $SRT$. Data points between $CFT$ and $SRT$ are normalized to the reaction time (duration between $CFT$ and $SRT$). $CFT$: clicker falling time; $SRT$: string release time; ML: medio-lateral; $t_0$: onset of CoP displacement away from the target.

Figure 3. Comparisons of the CoP ML displacement with all trials aligned with $t_0$ (onset of CoP displacement away from the target) between strategy (early/late – top panels) and expertise (elite/sub-elite – bottom panels). Left: timeseries mean and error cloud. Right: SPM timeseries. No temporal differences within strategy or expertise as there is no crossing over or under their respective $p$ value threshold (dotted lines).